\documentclass [6pt,a4paper]{article}
\usepackage{bbm}
\usepackage{amsfonts}
\usepackage{mathrsfs}
\usepackage {amssymb}
\usepackage {amsmath}
\usepackage{color}
\usepackage{amsthm}
\usepackage{cite}
\usepackage{latexsym}

\usepackage{lastpage}
\usepackage{epsfig}

\begin{document}
\newtheorem{lemma}{Lemma}[section]
\newtheorem{theorem}[lemma]{Theorem}
\newtheorem{example}[lemma]{Example}
\newtheorem{definition}[lemma]{Definition}
\newtheorem{proposition}[lemma]{Proposition}
\newtheorem{conjecture}[lemma]{Conjecture}
\newtheorem{remark}{Remark}
\begin{center}
\textbf{\large{Some new bounds on  LCD codes over finite fields}}\footnote { E-mail
addresses:
pbbmath@126.com(B.Pang), zhushixin@hfut.edu.cn(S.Zhu),  kxs6@sina.com(X.Kai). This research was supported by the National Natural Science Foundation of China under Grant Nos 61772168, 61572168 and 11501156, the Anhui Provincial Natural Science Foundation under Grant No 1508085SQA198.}\\
\end{center}

\begin{center}
{ Binbin Pang, Shixin Zhu,  Xiaoshan Kai}
\end{center}

\begin{center}
\textit{School of Mathematics, Hefei University of
Technology, Hefei 230009, Anhui, P.R.China }
\end{center}

\noindent\textbf{Abstract:}\   In this paper, we show that LCD codes are not equivalent to linear codes over small finite fields. The enumeration of binary optimal LCD codes is obtained. We also get the exact value of LD$(n,2)$ over $\mathbb{F}_3$ and $\mathbb{F}_4$. We study the bound of LCD codes over $\mathbb{F}_q$.         

\section{Introduction}
In this paper,  let $\mathbb{F}_q$ be a finite field with $q$ elements. The set of non-zero elements of $\mathbb{F}_q$ is denoted by $\mathbb{F}_q^\ast$. For any $x\in \mathbb{F}_{q^2}$, the conjugate of $x$ is defined as $\overline{x}=x^q$. A $k-$dimensional subspace $C$ of $\mathbb{F}_q^n$ is called  an $[n,k,d]$ linear code with minimum (Hamming) distance $d$. Given a linear code $C$ of length $n$ over $\mathbb{F}_q$ (resp. $\mathbb{F}_{q^2}$), its Euclidean dual code (resp. Hermitian dual code) is denoted by $C^\perp$ (resp. $C^{\perp_H}$). The codes  $C^\perp$ and $C^{\perp_H}$ are defined as follows
$$C^{\perp}=\{{\textbf{u}\in \mathbb{F}_q^n\mid \textbf{u}\cdot \textbf{c}=\textbf{0},\forall\  \textbf{c}\in C}\},$$
$$C^{\perp_H}=\{{\textbf{u}\in \mathbb{F}_q^n\mid \textbf{u}\cdot \overline{\textbf{c}}=\textbf{0},\forall\  \textbf{c}\in C}\}.$$

 A linear code has complementary dual (or LCD code for short) over $\mathbb{F}_q$  if $C\bigoplus C^\perp=\mathbb{F}_q^n$. The Euclidean (resp. Hermitian) hull of a linear code $C$ is defined to be $\textrm{Hull}_E(C)=C\cap C^{\perp}$ (resp. $\textrm{Hull}_H(C)=C\cap C^{\perp_H}$). A linear code over $\mathbb{F}_q$ is called a Euclidean (resp. Hermitian) LCD code if $\textrm{Hull}_E(C)=\{\mathbf{0}\}$ ($\textrm{Hull}_H(C)=\{\mathbf{0}\}$). In the later of this paper, Euclidean LCD code is abbreviated to LCD code if no special stated.

  In 1992, Massey first initiated LCD codes \cite{ref5},  and he also proved the existence of asymptotically good LCD codes. Sendrier showed that LCD codes meet the asymptotic Gilbert-Varshamov bound  over the finite fields \cite{ref6}. Yang and Maseey gave a necessary and sufficient condition for a  cyclic code to be LCD  over finite fields \cite{ref7}. After that, there are many literatures on the construction of LCD codes over finite fields \cite{ref8,ref9,ref10,ref11,ref12,ref13}.   What's more there are many LCD MDS code have been constructed by some scholars in \cite{ref15,ref16,ref17,ref18}. Carlet et al. solved the problem of the existence of $q$-ary $[n,k]$ LCD MDS codes for Euclidean case \cite{ref17}, they also introduced a general construction of LCD codes from any linear codes. Further more, they showed that any linear code over $\mathbb{F}_q\ (q > 3)$ is equivalent to an Euclidean LCD code and any linear code over $\mathbb{F}_{q^2}\ (q > 2)$ is equivalent to a Hermitian LCD code \cite{ref18}. Sok et al. proved the existence of optimal LCD codes over large finite fields \cite{ref19}. Liu et al. discussed the structure of LCD codes over finite chain rings\cite{ref14}.

  Recently, many researchers have  an interest in LCD codes over small finite fields\cite{ref20,ref21,ref22,ref23}. Galvez et al. gave bouds  on the minimum distances of binary LCD codes with fixed lengths and dimensions  on the dimensions of LCD codes with fixed lengths and minimum distances.\cite{ref21}.  Carlet et al. presented a new characterization of binary LCD codes in terms of their symplectic basis and solve a conjecture proposed by Galvez et al.\cite{ref22}.  Harada et al. studied binary LCD codes with the largest minimum weight among all binary LCD codes\cite{ref23}. Inspired by these latter works, we consider the bounds on LCD  codes over small finite fields.

  In this paper, we give some background and recall some basic results in Section 2. In Section 3, we show that LCD codes are not equivalent to linear codes over small finite fields. In Sections 4 , the enumeration of binary optimal LCD codes is obtained. In Sections 5 and 6, we get the exact value of LD$(n,2)$ over $\mathbb{F}_3$ and $\mathbb{F}_4$. In Section 7, we study the bound of LCD codes over $\mathbb{F}_q$.

\section{Preliminaries}

For any vector $\mathbf{a} = (a_1 , \cdots, a_n )\in \mathbb{F}_q^n$ and permutation $\sigma$ of $\{1,2, \cdots ,n\}$, we define $C_\mathbf{a}$ and $\sigma(C)$ as the following linear codes
$$C_\mathbf{a} = \{(a_1c_1, a_2c_2, \cdots , a_nc_n )\mid (c_1, c_2, \cdots ,c_n ) \in C\},$$
and
$$\sigma (C) = \{(c_{\sigma(1)}, c_{\sigma(2)}, \cdots ,c_{\sigma(n)}) \mid (c_1, c_2, \cdots ,c_n ) \in C\}.$$

Two codes $C$ and $C'$ in $\mathbb{F}_q^n$ are called equivalent if $C¡ä = \sigma(C_\mathbf{a} )$ for some permutation $\sigma$ of $\{1,2, \cdots ,n\}$ and $\mathbf{a}\in (\mathbb{F}_q^\ast)^n$. For a matrix $A$ over finite field, $A^T$ denotes the transposed matrix of $A$ and $\overline{A}$ denotes the conjugate of $A$. We assume that det$(A)$ denotes the determinant of A, where $A$ is a square matrix. Hamming weight vector  $\mathbf{a}$ is the number of nonzero $a_i$ , and denoted by wt$(\mathbf{a})$.

\begin{lemma}[see \cite{ref1}]\label{le:2.1}   If $G$ is a generator matrix for the $[n,k]$ linear code $C$, then $C$ is an Euclidean (resp. a Hermitian) LCD code if and only if, the $k\times k$ matrix $GG^T$ (resp. $G\overline{G}^T$) is nonsingular.
\end{lemma}

\begin{lemma} \label{le:2.2} Let $A$ be a $k\times n$ matrix. Let $a^1 , a^2, \cdots , a^n$ be the columns vectors of $A$ and $\widetilde{A}=[a^{\sigma(1)} , a^{\sigma(2)}, \cdots , a^{\sigma(n)}]$, where $\sigma$ is a permutation of $\{1, 2, \cdots, n\}$. Then det$(AA^T)$=det$(\widetilde{A}\widetilde{A}^T)$.
\end{lemma}
\proof  By the definition of $\widetilde{A}$, there are primary matrixes $Q_1, \cdots Q_s$ such that $\widetilde{A}=AQ_1\cdots Q_s$. We have $$\textrm{det}(\widetilde{A}\widetilde{A}^T)=\textrm{det}(AQ_1\cdots Q_sQ_1^{T}\cdots Q_s^{T}A^T)=\textrm{det}(AA^T).$$Then the proof is completed.\qed\\

The combinatorial functions LD$(n,k)$ and LD$(n,d)$ has been introduced and studied by Dougherty et al.  \cite{ref4} and Galvez et al. \cite{ref21}. The definitions of LD$(n,k)$ and LD$(n,d)$ as follows, we will use them frequently in the rest of this paper.

\begin{definition} \label{de:2.3} LD$(n,k):=max\{d\mid ~there ~exsits ~an ~[n,k,d]~ LCD~ code~ over~ \mathbb{F}_q\}$.
\end{definition}

\begin{definition} \label{de:2.4} LD$(n,d):=max\{k\mid ~there ~exsits ~an ~[n,k,d]~ LCD~ code~ over~ \mathbb{F}_q\}$.
\end{definition}

Let $C$ be an $[n,k,d]$ linear code over $\mathbb{F}_q$  and the matrix $G$ be the generator matrix of $C$. Then the size of $G$ is $k\times n$ and rank$(G)=k$. Let
\begin{gather*}
 G=\begin{bmatrix} a_{11} & a_{12} & \cdots & a_{1n}\\
                   a_{21} & a_{22} & \cdots & a_{2n}\\
                   \vdots & \vdots & \vdots & \vdots\\
                   a_{k1} & a_{k2} & \cdots & a_{kn}\\
\end{bmatrix},
\end{gather*}
where $a_{ij}\in \mathbb{F}_q$, for $1\leq i\leq k$ and $1\leq j\leq n$.

Any code over $\mathbb{F}_q$, we have the following inequality.
\begin{lemma} \label{le:2.6} $LD(n,k)\leq \lfloor \frac{(q-1)q^{k-1}n}{(q^k-1)}\rfloor$, for $1\leq k\leq n$.
\end{lemma}
\proof   From the Griesmer Bound  \cite{ref3}, for any $q$-ary linear $[n,k,d]$ code, we have $$n\geq\sum_{i=0}^{k-1}\lceil\frac{d}{q^i}\rceil.$$ we have $n\geq \frac{dq(q^k-1)}{(q-1)q^k}$. Hence $$d\leq \lfloor \frac{(q-1)q^kn}{q(q^k-1)}\rfloor.$$ Therefore any $[n,k,d]$ LCD code must satisfy this inequality.\qed

\begin{lemma} \label{le:2.7}
Let  $n$ and $k$ are positive integers, $k>0$, then LD$(n + 1,k)\geq $LD$(n,k)$
\end{lemma}
\proof  Let G be a generator matrix of an $[n, k, d]$ LCD code $C$ over $\mathbb{F}_q$. Then $C'$ with the generator matrix $G'=[G ~\mathbf{0}]$ is an LCD code since det$(G'G'^T)$=det$(GG^T)\neq0$. Note that $C'$ is an $[n+1, k, d]$ code. This completes the proof.\qed

\section{LCD codes are not equivalent to  linear codes over small finite fields}
In this section, we investigate  the relationship between linear codes and LCD codes over small finite fields. In \cite{ref1}, Carlet et al. showed that an $[n,k,d]$ linear Euclidean LCD code over $\mathbb{F}_q$ with $q > 3$ exists if and only if  there is an $[n,k,d]$ linear code over $\mathbb{F}_q$ and an $[n,k,d]$ linear
Hermitian LCD code over $\mathbb{F}_{q^2}$ with $q > 2$ exists if and only if there is an $[n,k,d]$ linear code over $\mathbb{F}_{q^2}$. Now we proved that this result is not true in the small finite fields, such as $\mathbb{F}_2$, $\mathbb{F}_3$ and $\mathbb{F}_4$.\\

\begin{theorem}  Let $C$ be a linear code over $\mathbb{F}_2$ with generator matrix $G$ and assume $C$ is not an Euclidean LCD code.  Then $C$ is not equivalent to any Euclidean LCD codes over $\mathbb{F}_2$.
\end{theorem}
\proof  Assume $C$ is equivalent to a linear code $\widetilde{C}=\sigma(C_\mathbf{a})$, where $\sigma$ is a permutation of $\{1, 2, \cdots, n\}$ and $\mathbf{a}\in (\mathbb{F}_2^\ast)^n$. It is obvious that $\mathbf{a}=(1,1,\cdots,1)$. Let $G$ and $\widetilde{G}$ are the generator matrixes of $C$ and $C'$, respectively. It is easy to know det($GG^T$)=0, then det$(\widetilde{G}\widetilde{G}^T)=0$ from Lemma \ref{le:2.2}. We show that linear code $\widetilde{C}$ is not to be Euclidean LCD.\qed

\begin{theorem}  Let $C$ be a linear code over $\mathbb{F}_3$ with generator matrix $G$ and assume $C$ is not an  Euclidean LCD code.  Then $C$ is not equivalent to any Euclidean LCD codes over $\mathbb{F}_3$.
\end{theorem}

\proof Assume $C$ is equivalent to a linear code $\widetilde{C}=\sigma(C_\mathbf{a})$, where $\sigma$ is a permutation of $\{1, 2, \cdots, n\}$ and $\mathbf{a} = (a_1 , \cdots, a_n )\in (\mathbb{F}_3^\ast)^n$. Let $G$, $G_\mathbf{a}$ and $\widetilde{G}$ are the generator matrixes of $C$, $C_\mathbf{a}$ and $C'$, respectively.  The $G_\mathbf{a}$ is obtained from $G$ by multiplying its $j-$th column by $a_j$ for $j\in\{1,2, \cdots n\}$, then we have $G_\mathbf{a}G_\mathbf{a}^T=GG^T$ by simple matrix operations. It is easy to know det($GG^T$)=0, then det$(\widetilde{G}\widetilde{G}^T)=\textrm{det}(G_\mathbf{a}G_\mathbf{a}^T)=\textrm{det}(GG^T)=0$ from Lemma \ref{le:2.2}. We show that linear code $\widetilde{C}$ is not to be Euclidean LCD.\qed

\begin{theorem}  Let $C$ be a linear code over $\mathbb{F}_4$ with generator matrix $G$ and assume $C$ is not a Hermitian LCD code.  Then $C$ is not equivalent to any Hermitian LCD codes over $\mathbb{F}_4$.
\end{theorem}

\proof Assume $C$ is equivalent to a linear code $\widetilde{C}=\sigma(C_\mathbf{a})$, where $\sigma$ is a permutation of $\{1, 2, \cdots, n\}$ and $\mathbf{a} = (a_1 , \cdots, a_n )\in (\mathbb{F}_4^\ast)^n$. Let $G$, $G_\mathbf{a}$ and $\widetilde{G}$ are the generator matrixes of $C$, $C_\mathbf{a}$ and $C'$, respectively.  The $G_\mathbf{a}$ is obtained from $G$ by multiplying its $j-$th column by $a_j$ for $j\in\{1,2, \cdots n\}$ and $2^2-1=2+1$, then we have $G_\mathbf{a}\overline{G}_\mathbf{a}^T=G\overline{G}^T$ by simple matrix operations. It is easy to know det($G\overline{G}^T$)=0, then det$(\widetilde{G}\overline{\widetilde{G}}^T)=\textrm{det}(G_\mathbf{a}\overline{G}_\mathbf{a}^T)=\textrm{det}(G\overline{G}^T)=0$ from Lemma \ref{le:2.2}. We show that linear code $\widetilde{C}$ is not to be Hermitian LCD. \qed

Thus in the later section, we only consider LCD codes over $\mathbb{F}_2$, $\mathbb{F}_3$ and $\mathbb{F}_4$.
\section{The enumeration of $[n,2,d]$ binary optimal LCD  codes}

In this section we consider binary codes. Recently, Galvez et al. \cite{ref4} obtain the exact values of LD$(n,k)$ for $k=2$ and arbitrary $n$.  By Theorem 1 in \cite{ref4}, we know that there exist LCD codes with LD$(n,2)=\lfloor\frac{2n}{3}\rfloor$ only for $n\equiv1, \pm2, 3~(mod~6)$. An $[n,k,d]$ linear code is optimal if the minimum distance achieve the Gresmer Bound. In this section, we will give the enumeration of $[n,2,d]$ binary optimal LCD  codes for  $n\equiv1, \pm2, 3~(mod~6)$, where $d=\lfloor\frac{2n}{3}\rfloor$.

 An $[n,k]$ linear code $C$ over $\mathbb{F}_2$ with generator matrix $G$, Let
\begin{gather*}
 G=\begin{bmatrix} a_{11} & a_{12} & \cdots & a_{1n}\\
                   a_{21} & a_{22} & \cdots & a_{2n}\\
\end{bmatrix}.
\end{gather*}
Let $\alpha_1=[a_{11} , a_{12} , \cdots, a_{1n}]$, $\alpha_2=[a_{21}, a_{22},\cdots, a_{2n}]$, then
\begin{gather*}
 G=\begin{bmatrix}\alpha_1\\
                   \alpha_2\\
\end{bmatrix}.
\end{gather*}
Let $\beta_l=[a_{1l}, a_{2l}]^T$ for $1\leq l\leq n$, then $G=[\beta_1, \beta_2, \cdots, \beta_n]$.

The following definition will be frequently in this section.
\begin{definition} \label{de:4.1} $S_{ij}:=\mid\{i\mid [i, j]^T=\beta_l\mid for~ 1\leq l\leq n\}\mid$, for $i,j\in \mathbb{F}_2$.
\end{definition}

From this definition and notation given above, we have $C=\{0, \alpha_1, \alpha_2,  \alpha_1+\alpha_2, \}$. It is easy to know $$wt(\alpha_1)=S_{10}+S_{11}, wt(\alpha_2)=S_{01}+S_{11}, wt(\alpha_1+\alpha_2)=S_{10}+S_{01}.$$
We also have
\begin{gather*}
 GG^T=\begin{bmatrix} S_{10}+S_{11} &  S_{11}\\
                      S_{11}        &  S_{01}+S_{11} \\
\end{bmatrix}.
\end{gather*}

Based on the notation given above, we can obtain the following theorems.
\begin{theorem} \label{th:4.2}
Up to equivalence, the number of $[n,2,d]$ binary optimal LCD codes is $2$, for $n\equiv1, \pm2~(mod~6)$.
\end{theorem}
\proof  (i) Let $n\equiv1 ~(mod~6)$, i.e., $n=6t+1$, for some positive integer $t$. Let the code with generator matrix $G$, let min$\{wt(\alpha_1), wt(\alpha_2),  wt(\alpha_1+\alpha_2)\}\geq\lfloor\frac{2(6t+1)}{3}\rfloor=4t$, then $(S_{01}, S_{10}, S_{11})\in T$, where $T=\{(2t-1, 2t+1, 2t+1,), (2t, 2t, 2t+1,), (2t, 2t+1, 2t,), (2t+1, 2t-1, 2t+1,), (2t+1, 2t, 2t,), (2t+1, 2t+1, 2t-1,)\}$. If $(S_{01}, S_{10}, S_{11})\in T_1$, where $T_1=\{ (2t-1, 2t+1, 2t+1,), (2t+1, 2t-1, 2t+1,),  (2t+1, 2t+1, 2t-1,)\}$. Note that the matrix $G$ of those codes always satisfy det$(GG^T)\neq0$. Therefor those codes are  LCD code. But the code generator by $(S_{01}, S_{10}, S_{11})=(2t-1, 2t+1, 2t+1,)$ is equivalent to the code generator by $(S_{01}, S_{10}, S_{11})=(2t+1, 2t-1, 2t+1,)$. If $(S_{01}, S_{10}, S_{11})\in T_2$, where $T_2=T\setminus T_1$. Note that the matrix $G$ of those codes always satisfy det$(GG^T)=0$. Therefor those codes are not  LCD code. Hence, there are only two binary optimal LCD codes.

(ii) Let $n\equiv-2 ~(mod~6)$, i.e., $n=6t-2$, for some positive integer $t$. The proof is similar to (i), we omit detail here.We obtain $(S_{01}, S_{10}, S_{11})\in T$, where $T=\{(2t-2, 2t, 2t,), (2t-1, 2t-1, 2t,), (2t-1, 2t, 2t-1,), (2t, 2t-2, 2t,), (2t, 2t-1, 2t-1,), (2t, 2t, 2t-2,)\}$. If $(S_{01}, S_{10}, S_{11})\in T_1$, where $T_1=\{ (2t-1, 2t-1, 2t,), (2t-1, 2t, 2t-1,),  (2t, 2t-1, 2t-1,)\}$. Note that the matrix $G$ of those codes always satisfy det$(GG^T)\neq0$. Therefor those codes are  LCD code. But the code generator by $(S_{01}, S_{10}, S_{11})=(2t-1, 2t, 2t-1,)$ is equivalent to the code generator by $(S_{01}, S_{10}, S_{11})=(2t, 2t-1, 2t-1,)$. If $(S_{01}, S_{10}, S_{11})\in T_2$, where $T_2=T\setminus T_1$. Note that the matrix $G$ of those codes always satisfy det$(GG^T)=0$. Therefor those codes are not  LCD code. Hence, there are only two binary optimal LCD codes.

(iii) Let $n\equiv2 ~(mod~6)$, i.e., $n=6t+2$, for some positive integer $t$. The proof is similar to (i), we omit detail here. We obtain $T=\{(2t, 2t+1, 2t+1,), (2t+1, 2t, 2t+1,), (2t+1, 2t+1, 2t,)$. If $(S_{01}, S_{10}, S_{11})\in T$,  note that the matrix $G$ of those codes always satisfy det$(GG^T)\neq0$. Therefor those codes are  LCD code. But the code generator by $(S_{01}, S_{10}, S_{11})=(2t, 2t+1, 2t+1,)$ is equivalent to the code generator by $(S_{01}, S_{10}, S_{11})=(2t+1, 2t, 2t+1,)$.  Hence, there are only two binary optimal LCD codes.\qed

\begin{theorem} \label{th:4.3}
Up to equivalence, the number of $[n,2,d]$ binary optimal LCD codes is $1$, for $n\equiv3~(mod~6)$.
\end{theorem}
\proof   Let $n\equiv3 ~(mod~6)$, i.e., $n=6t+3$, for some positive integer $t$. Let the code with generator matrix $G$, let min$\{wt(\alpha_1), wt(\alpha_2),  wt(\alpha_1+\alpha_2)\}\geq\lfloor\frac{2(6t+3)}{3}\rfloor=4t+2$, then $(S_{01}, S_{10}, S_{11})\in T$, where $T=\{(2t-1, 2t+2, 2t+2,), (2t, 2t+1, 2t+2,), (2t, 2t+2, 2t+1,), (2t+1, 2t, 2t+2,), (2t+1, 2t+1, 2t+1,), (2t+1, 2t+2, 2t,), (2t+2, 2t-1, 2t+2,), (2t+2, 2t, 2t+1,), (2t+2, 2t+1, 2t,), (2t+2, 2t+2, 2t-1,)\}$. If $(S_{01}, S_{10}, S_{11})\in T_1$, where $T_1=\{ (2t+1, 2t+1, 2t+1,)\}$. Note that the generator matrix $G$ of this code satisfy $GG^T=\left[
\begin{array}{ccc}
0 & 1 \\
1 & 0 \\
\end{array}
\right]$ i.e., det$(GG^T)=1\neq0$. Therefor this code is an  LCD code.  If $(S_{01}, S_{10}, S_{11})\in T_2$, where $T_2=T\setminus T_1$. Note that the matrix $G$ of those codes always satisfy det$(GG^T)=0$. Therefor those codes are not  LCD code. Hence, there are only one binary optimal LCD code. \qed

\begin{theorem} \label{th:6.2} The let $n$ be a positive integer, then

(1) Suppose that  $n$ is even and $i\geq0$. If $n>6i+3$, then LK$(n,n-2i-1)$=1.

(2) Suppose that  $n$ is odd and $i\geq0$. If $n\geq6i$, then LK$(n,n-2i)$=1.
\end{theorem}
\proof  (1) Let $C$ be an $[n, k, n-2i-1]$ LCD code over $\mathbb{F}_2$ with generator matrix G. If $k\geq2$, there are an $[n, 2, n-2i-1]$ LCD code by Theorem 3.4 in \cite{ref22}. From the Griesmer bound, we have $n\leq 6i+3$. This is imply that there is no $[n, 2, n-2i-1]$ code when $n>6i+3$.  That is, there is no such
an LCD code.

If $k=1$, because the minimum distance $n-2i-1$ is odd, thus we get $GG^T=1$. There is $[n, 1, n-2i-1]$ LCD code by Lemma \ref{le:2.1}.

 (2) Let $C$ be an $[n, k, n-2i]$ LCD code over $\mathbb{F}_2$ with generator matrix G. If $k\geq2$, there are an $[n, 2, n-2i]$ LCD code by Theorem 3.4 in \cite{ref22}. From the Griesmer bound, we have $n\leq 6i$. Since LD$[6i,2)=4i-1$ from Theorem 2 in \cite{ref21}. Thus there is no $[6i, 2, 4i]$. This is imply that there is no $[n, 2, n-2i-1]$ code when $n\geq6i$.  That is, there is no such an LCD code.

If $k=1$, because the minimum distance $n-2i$ is odd, thus we get $GG^T=1$. There is $[n, 1, n-2i-1]$ LCD code by Lemma \ref{le:2.1}. \qed

\section{The exact value of LD$(n,2)$ over $\mathbb{F}_3$}

In this section, we consider ternary codes and give the exact value of LD$(n,2)$ over $\mathbb{F}_3$ for $k = 2$ and arbitrary $n$.

An $[n,k]$ linear code $C$ over $\mathbb{F}_3$ with generator matrix $G$, Let
\begin{gather*}
 G=\begin{bmatrix} a_{11} & a_{12} & \cdots & a_{1n}\\
                   a_{21} & a_{22} & \cdots & a_{2n}\\
\end{bmatrix}.
\end{gather*}
Let $\alpha_1=[a_{11} , a_{12} , \cdots, a_{1n}]$, $\alpha_2=[a_{21}, a_{22},\cdots, a_{2n}]$, then
\begin{gather*}
 G=\begin{bmatrix}\alpha_1\\
                   \alpha_2\\
\end{bmatrix}.
\end{gather*}
Let $\beta_l=[a_{1l}, a_{2l}]^T$ for $1\leq l\leq n$, then $G=[\beta_1, \beta_2, \cdots, \beta_n]$.

The following definition will be frequently in this section.
\begin{definition} \label{de:5.1} $S_{ij}:=\mid\{i\mid [i, j]^T=\beta_l\mid for~ 1\leq l\leq n\}\mid$, for $i,j\in \mathbb{F}_3$.
\end{definition}

From this definition and notation given above, we have $C=\{0, \alpha_1, \alpha_2, 2\alpha_1, 2\alpha_2,  \alpha_1+\alpha_2, \alpha_1+2\alpha_2, 2\alpha_1+\alpha_2, 2\alpha_1+2\alpha_2\}$. It is easy to know $$wt(\alpha_1)=wt(2\alpha_1)=\sum_{i=1}^2\sum_{j=0}^2S_{ij}, wt(\alpha_2)=wt(2\alpha_2)=\sum_{i=0}^2\sum_{j=1}^2S_{ij},$$ $$wt(\alpha_1+\alpha_2)=wt(2\alpha_1+2\alpha_2)=S_{01}+S_{02}+S_{10}+S_{11}+S_{20}+S_{22},$$ $$wt(\alpha_1+2\alpha_2)= wt(2\alpha_1+\alpha_2)=S_{01}+S_{02}+S_{10}+S_{12}+S_{20}+S_{21}.$$
We also have
\begin{gather*}
 GG^T=\begin{bmatrix} \sum_{i=1}^2\sum_{j=0}^2i^2S_{ij} &  \sum_{i=1}^2\sum_{j=1}^2ijS_{ij}\\
                      \sum_{i=1}^2\sum_{j=1}^2ijS_{ij} &  \sum_{i=0}^2\sum_{j=1}^2j^2S_{ij} \\
\end{bmatrix}.
\end{gather*}

Based on the notation given above, we can obtain the following theorems.

\begin{theorem} \label{th:5.2}
$LD(n,2)\leq \lfloor\frac{3n}{4}\rfloor$ for $n\geq2$
\end{theorem}
\proof   From the   Lemma \ref{le:2.6}, let $q=3,$  and $k=2$, we get this inequality.\qed

\begin{theorem} \label{th:5.3} Let $n\geq2$. Then $LD(n,2)= \lfloor\frac{3n}{4}\rfloor$ for $n\equiv1, 2 ~(mod~4)$.
\end{theorem}
\proof   Let linear code $C$ generate by $G$ give above,  we only need to show the existence of LCD code with minimum distance $d=\lfloor\frac{3n}{4}\rfloor$.

(i) Let $n\equiv1 ~(mod~4)$, i.e., $n=4t+1$, for some positive integer $t$. If $t$ is an odd integer, let the code with generator matrix $G$, let $S_{01}=S_{02}=S_{10}=S_{12}=S_{21}=\frac{t+1}2$ and $S_{11}=S_{20}=S_{22}=\frac{t-1}2$. Note that this code has minimum distance $3t=\lfloor\frac{3(4t+1)}{4}\rfloor$ and
$GG^T=\left[
\begin{array}{ccc}
 0 & 1 \\
 1 & 1 \\
\end{array}
\right]$ i.e., det$(GG^T)=2\neq0$. Therefor this code is an LCD code. If $t$ is an even integer, let the code with generator matrix $G$, let $S_{01}=S_{02}=S_{10}=S_{11}=S_{12}=\frac{t}2$, $S_{21}=\frac{t}2-1$ and $S_{20}=S_{22}=\frac{t}2+1$, Note that this code has minimum distance $3t=\lfloor\frac{3(4t+1)}{4}\rfloor$ and
$GG^T=\left[
\begin{array}{ccc}
 1 & 2 \\
 2 & 0 \\
\end{array}
\right]$ i.e., det$(GG^T)=2\neq0$. Therefor this code is an LCD code.

(ii) Let $n\equiv2 ~(mod~4)$, i.e., $n=4t+2$, for some positive integer $t$. If $t$ is an odd integer, let the code with generator matrix $G$, let $S_{01}=S_{02}=S_{11}=S_{12}=S_{20}=S_{22}=\frac{t+1}2$ and $S_{10}=S_{21}=\frac{t-1}2$, Note that this code has minimum distance $3t+1=\lfloor\frac{3(4t+2)}{4}\rfloor$ and
$GG^T=\left[
\begin{array}{ccc}
 1 & 1 \\
 1 & 2 \\
\end{array}
\right]$ i.e., det$(GG^T)=1\neq0$. Therefor this code is an LCD code. If $t$ is an even integer, let the code with generator matrix $G$, let $S_{01}=S_{02}=S_{12}=S_{20}=S_{21}=S_{22}=\frac{t}2$ and $S_{10}=S_{11}=\frac{t}2+1$. Note that this code has minimum distance $3t+1=\lfloor\frac{3(4t+2)}{4}\rfloor$ and
$GG^T=\left[
\begin{array}{ccc}
 2 & 1 \\
 1 & 1 \\
\end{array}
\right]$ i.e., det$(GG^T)=1\neq0$. Therefor this code is an LCD code.\qed

\begin{theorem} \label{th:5.4} Let $n\geq2$. Then $LD(n,2)= \lfloor\frac{3n}{4}\rfloor-1$ for $n\equiv0, 3 ~(mod~4)$.
\end{theorem}
\proof   Let linear code $C$ generate by $G$ give above,  we will show there is no LCD code with minimum distance $d=\lfloor\frac{3n}{4}\rfloor$.

(i) Let $n\equiv0 ~(mod~4)$, i.e., $n=4t$, for some positive integer $t$. Let the code with generator matrix $G$, let min$\{wt(\alpha_1), wt(\alpha_2),  wt(\alpha_1+\alpha_2), wt(\alpha_1+2\alpha_2) \}\geq\lfloor\frac{3(4t)}{4}\rfloor=3t$, then, $wt(\alpha_1)=wt(\alpha_2)= wt(\alpha_1+\alpha_2)=wt(\alpha_1+2\alpha_2)=3t$ and $S_{01}+S_{02}=S_{10}+S_{20}=S_{11}+S_{22}=S_{12}+S_{21}=t$. Note that
$GG^T=\left[
\begin{array}{ccc}
0 & 0 \\
0 & 0 \\
\end{array}
\right]$ i.e., det$(GG^T)=0$. Therefor those codes are not LCD codes. Furthermore,  if $t$ is an odd integer, let the code with generator matrix $G$, let $S_{01}=S_{02}=S_{11}=S_{12}=\frac{t-1}2$ and $S_{10}=S_{20}=S_{21}=S_{22}=\frac{t+1}2$. Note that this code has minimum distance $3t-1=\lfloor\frac{3(4t)}{4}\rfloor-1$ and
$GG^T=\left[
\begin{array}{ccc}
 1 & 0 \\
 0 & 2 \\
\end{array}
\right]$ i.e., det$(GG^T)=2\neq0$. Therefor this code is an LCD code. If $t$ is an even integer, let the code with generator matrix $G$, let $S_{01}=S_{02}=S_{10}=S_{11}=S_{12}=S_{20}=\frac{t}2$, $S_{21}=\frac{t}2-1$ and $S_{22}=\frac{t}2+1$. Note that this code has minimum distance $3t-1=\lfloor\frac{3(4t)}{4}\rfloor-1$ and
$GG^T=\left[
\begin{array}{ccc}
 0 & 2 \\
 2 & 0 \\
\end{array}
\right]$ i.e., det$(GG^T)=2\neq0$. Therefor this code is an LCD code.

(ii) Let $n\equiv3 ~(mod~4)$, i.e., $n=4t+3$, for some positive integer $t$. Let the code with generator matrix $G$, let min$\{wt(\alpha_1), wt(\alpha_2),  wt(\alpha_1+\alpha_2), wt(\alpha_1+2\alpha_2) \}\geq\lfloor\frac{3(4t+3)}{4}\rfloor=3t+2$, then, $wt(\alpha_1), wt(\alpha_2), wt(\alpha_1+\alpha_2), wt(\alpha_1+2\alpha_2)\in\{3t+2, 3t+3\}$. (1). Let $wt(\alpha_1)=wt(\alpha_2)=3t+3$, then $S_{01}+S_{02}=S_{10}+S_{20}=t$, if $wt(\alpha_1+\alpha_2)=3t+3$, we have $S_{12}+S_{21}=t$, then $S_{11}+S_{22}=t+3$, we obtain $wt(\alpha_1+2\alpha_2)=3t$, which is impossible. If $wt(\alpha_1+\alpha_2)=3t+2$, we have $S_{12}+S_{21}=t+1$, then $S_{11}+S_{22}=t+2$, we obtain $wt(\alpha_1+2\alpha_2)=3t+1$, which is also impossible. (2). Let $wt(\alpha_1)=3t+2$, $wt(\alpha_2)=3t+3$, then $S_{01}+S_{02}=t+1$, $S_{10}+S_{20}=t$, if $wt(\alpha_1+\alpha_2)=3t+3$, we have $S_{12}+S_{21}=t$, then $S_{11}+S_{22}=t+2$, we obtain $wt(\alpha_1+2\alpha_2)=3t+1$, which is impossible. If $wt(\alpha_1+\alpha_2)=3t+2$, we have $S_{12}+S_{21}=t+1$, then $S_{11}+S_{22}=t+1$. Note that
$GG^T=\left[
\begin{array}{ccc}
2 & 0 \\
0 & 0 \\
\end{array}
\right]$ i.e., det$(GG^T)=0$. Therefor this code is not an LCD code. (3). Let $wt(\alpha_1)=3t+3$, $wt(\alpha_2)=3t+2$. It is similar to (2). We get
$GG^T=\left[
\begin{array}{ccc}
0 & 0 \\
0 & 2 \\
\end{array}
\right]$ i.e., det$(GG^T)=0$. Therefor this code is not an LCD code. (4). Let $wt(\alpha_1)=wt(\alpha_2)=3t+2$, then $S_{01}+S_{02}=S_{10}+S_{20}=t+1$, if $wt(\alpha_1+\alpha_2)=3t+3$, we have $S_{12}+S_{21}=t$, then $S_{11}+S_{22}=t+1$. Note that
$GG^T=\left[
\begin{array}{ccc}
2 & 1 \\
1 & 2 \\
\end{array}
\right]$ i.e., det$(GG^T)=0$. Therefor this code is not an LCD code. If $wt(\alpha_1+\alpha_2)=3t+2$, we have $S_{11}+S_{22}=t+1$, then $S_{12}+S_{21}=t$. Note that
$GG^T=\left[
\begin{array}{ccc}
2 & 2 \\
2 & 2 \\
\end{array}
\right]$ i.e., det$(GG^T)=0$. Therefor this code is not an LCD code. Furthermore,  if $t$ is an odd integer, let the code with generator matrix $G$, let $S_{01}=S_{02}=S_{10}=S_{11}=S_{12}=S_{20}=\frac{t+1}2$, $S_{21}=\frac{t+3}2$ and $S_{22}=\frac{t-3}2$. Note that this code has minimum distance $3t+1=\lfloor\frac{3(4t+3)}{4}\rfloor-1$ and
$GG^T=\left[
\begin{array}{ccc}
 2 & 0 \\
 0 & 2 \\
\end{array}
\right]$ i.e., det$(GG^T)=1\neq0$. Therefor this code is an LCD code. If $t$ is an even integer, let the code with generator matrix $G$, let $S_{01}=S_{02}=S_{10}=S_{11}=S_{22}=\frac{t}2$ and $S_{12}=S_{20}=S_{21}=\frac{t}2+1$. Note that this code has minimum distance $3t+1=\lfloor\frac{3(4t+3)}{4}\rfloor-1$ and
$GG^T=\left[
\begin{array}{ccc}
 0 & 1 \\
 1 & 2 \\
\end{array}
\right]$ i.e., det$(GG^T)=2\neq0$. Therefor this code is an LCD code. This completes the proof.\qed

\section{The exact value of LD$(n,2)$ over $\mathbb{F}_4$}

In this section, we consider quaternary codes and give the exact value of LD$(n,2)$ over $\mathbb{F}_4$ for $k = 2$ and arbitrary $n$. Let $\xi$ be a primitive element of $\mathbb{F}_4$, i.e., $\mathbb{F}_4=\langle\xi\rangle\bigcup\{0\}$.

An $[n,k]$ linear code $C$ over $\mathbb{F}_4$ with generator matrix $G$. Let
\begin{gather*}
 G=\begin{bmatrix} a_{11} & a_{12} & \cdots & a_{1n}\\
                   a_{21} & a_{22} & \cdots & a_{2n}\\
\end{bmatrix}.
\end{gather*}
Let $\alpha_1=[a_{11} , a_{12} , \cdots, a_{1n}]$, $\alpha_2=[a_{21}, a_{22},\cdots, a_{2n}]$, then
\begin{gather*}
 G=\begin{bmatrix}\alpha_1\\
                   \alpha_2\\
\end{bmatrix}.
\end{gather*}
Let $\beta_l=[a_{1l}, a_{2l}]^T$ for $1\leq l\leq n$, then $G=[\beta_1, \beta_2, \cdots, \beta_n]$.

The following definition will be frequently in this section.
\begin{definition} \label{de:5.1} $S_{ij}:=\mid\{i\mid [i, j]^T=\beta_l\mid for~ 1\leq l\leq n\}\mid$, for $i,j\in \mathbb{F}_4$.
\end{definition}

From this definition and notation given above, we have $C=\{0, \alpha_1, \alpha_2, \xi\alpha_1, \xi\alpha_2, \xi^2\alpha_1, \xi^2\alpha_2, \alpha_1+\alpha_2, \xi\alpha_1+\xi\alpha_2, \xi^2\alpha_1+\xi^2\alpha_2, \alpha_1+\xi\alpha_2, \xi\alpha_1+\xi^2\alpha_2, \xi^2\alpha_1+\alpha_2, \alpha_1+\xi^2\alpha_2, \xi\alpha_1+\alpha_2, \xi^2\alpha_1+\xi\alpha_2\}$. It is easy to know $$wt(\alpha_1)=wt(\xi\alpha_1)=wt(\xi^2\alpha_1)=\sum_{i\in\mathbb{F}_4^*}\sum_{j\in\mathbb{F}_4}S_{ij}, wt(\alpha_2)=wt(\xi\alpha_2)=wt(\xi^2\alpha_2)=\sum_{i\in\mathbb{F}_4}\sum_{j\in\mathbb{F}_4^*}S_{ij},$$
$$wt(\alpha_1+\alpha_2)=wt(\xi\alpha_1+\xi\alpha_2)=wt(\xi^2\alpha_1+\xi^2\alpha_2)=\sum_{i\in\mathbb{F}_4}\sum_{j\in\mathbb{F}_4\atop i\neq j}S_{ij},$$
$$wt(\alpha_1+\xi\alpha_2)=wt(\xi\alpha_1+\xi^2\alpha_2)=wt(\xi^2\alpha_1+\alpha_2)=\sum_{i\in\mathbb{F}_4}\sum_{j\in\mathbb{F}_4\atop i\neq\xi j}S_{ij},$$
$$wt(\alpha_1+\xi^2\alpha_2)= wt(\xi\alpha_1+\alpha_2)= wt(\xi^2\alpha_1+\xi\alpha_2)=\sum_{i\in\mathbb{F}_4}\sum_{j\in\mathbb{F}_4\atop i\neq\xi^2 j}S_{ij}.$$
Let $y_0=S_{11}+S_{\xi\xi}+S_{\xi^2\xi^2}$, $y_1=S_{1\xi^2}+S_{\xi1}+S_{\xi^2\xi}$, $y_2=S_{1\xi^2}+S_{\xi1}+S_{\xi^21\xi}$, $y_{11}=\sum_{i\in\mathbb{F}_4^*}\sum_{j\in\mathbb{F}_4}S_{ij}$, $y_{22}=\sum_{i\in\mathbb{F}_4}\sum_{j\in\mathbb{F}_4^*}S_{ij}$. We also have
\begin{gather*}
 G\overline{G}^T=\begin{bmatrix} y_{11} &  y_0+y_1\xi+y_2\xi^2\\
                      y_0+y_2\xi+y_1\xi^2 &  y_{22} \\
\end{bmatrix}.
\end{gather*}

Based on the notation given above, we can obtain the following theorems.

\begin{theorem} \label{th:5.2}
$LD(n,2)\leq \lfloor\frac{4n}{5}\rfloor$ for $n\geq2$
\end{theorem}
\proof   From the   Lemma \ref{le:2.6}, let $q=4,$  and $k=2$, we get this inequality.\qed

\begin{theorem} \label{th:5.31} Let $n\geq2$. Then $LD(n,2)= \lfloor\frac{4n}{5}\rfloor$ for $n\equiv1, 2, 3 ~(mod~5)$.
\end{theorem}
\proof   It is similar to the Theorem \ref{th:5.3}, so we omit it.\qed

\begin{theorem} \label{th:5.41} Let $n\geq2$. Then $LD(n,2)= \lfloor\frac{4n}{5}\rfloor-1$ for $n\equiv0, 4 ~(mod~5)$.
\end{theorem}
\proof   It is similar to the Theorem \ref{th:5.4}, so we omit it.\qed

\section{Bound of $[n,k]$ LCD codes over $\mathbb{F}_q$}
In this section, we consider LCD codes over $\mathbb{F}_q$, where $q\geq3$. We get a relation between  LD$(n,k)$ and LD$(n,k-1)$.
Let $C$ be an $[n,k,d]$ linear code over $\mathbb{F}_q$  and the matrix $G$ be the generator matrix of $C$. Then the size of $G$ is $k\times n$ and rank$(G)=k$. Let
\begin{gather*}
 G=\begin{bmatrix} \alpha_{1} \\
                   \alpha_{2} \\
                   \vdots \\
                   \alpha_{k} \\
\end{bmatrix},
\end{gather*}
where $\alpha_{i}\in \mathbb{F}_q^n$, for $1\leq i\leq k$.

\begin{lemma} \label{le:6.1} Let $C$ be an $[n,k]$  LCD code over $\mathbb{F}_q$, where $k<n$. Then there exist a nonzero codeword $\beta\in C^\perp$ such that $\beta\cdot\beta=b\neq0$.
\end{lemma}
\proof  If $\beta\neq0$ and $\beta\cdot\beta=0$. For any  $\gamma\in C^\perp$, $(\beta+\gamma)\cdot(\beta+\gamma)=0$,  then  $\beta\in C$. This is impossible.\qed

\begin{theorem} \label{th:6.1} If $1\leq k\leq n$, then LD$(n,k)\leq$LD$(n,k-1)$.
\end{theorem}
\proof Let $C$ be an $[n,k-1]$ LCD code over $\mathbb{F}_q$ with generator matrix  $G=\left[
\begin{array}{ccc}
 \alpha_{1} \\
                   \alpha_{2} \\
                   \vdots \\
                   \alpha_{k-1} \\
\end{array}
\right]$, where $\alpha_{i}\in \mathbb{F}_q^n$, for $1\leq i\leq k-1$. Let $A=GG^T$, we have det$(A)\neq0$ by  Lemma \ref{le:2.1}. We let $C'$ be a code over $\mathbb{F}_q$ with generator matrix $G'=\left[
\begin{array}{ccc}
                   G \\
                  \beta \\
\end{array}
\right]=\left[
\begin{array}{ccc}
                  \alpha_{1} \\
                   \alpha_{2} \\
                   \vdots \\
                   \alpha_{k-1} \\
                       \beta \\
\end{array}
\right]$, where $\beta \in C^\perp$, $\beta\cdot\beta=b\neq0$ by Lemma \ref{le:6.1}. Let $A'=G'G'^T=\left[
\begin{array}{ccc}
                  A & \mathbf{0} \\
                  \mathbf{0} & b \\
\end{array}
\right]$, we have det$(A')=b\cdot$det$(A)\neq0$. Then $C'$ is an $[n,k]$ LCD code over $\mathbb{F}_q$ by  Lemma \ref{le:2.1}. It is easy to know $C\subseteq C'$, so we have $d(C)\geq d(C')$. Then LD$(n,k)\leq$LD$(n,k-1)$.  \qed

\section{Conclusion}
In this paper, we show that LCD codes are not equivalent to linear codes over small finite fields. The enumeration of binary optimal LCD codes is obtained. We also get the exact value of LD$(n,2)$ over $\mathbb{F}_3$ and $\mathbb{F}_4$. The techniques presented in this paper can be used for bound of minimum distance with larger dimensions. We study the bound of LCD codes over $\mathbb{F}_q$.


\end{document}